\documentclass[12pt, preprint]{aastex}
\received{ }
\revised{ }
\accepted{ }
\cpright{AAS}{2000}
\shorttitle{Cool ISM in Ellipticals}
\shortauthors{Welch, Sage \& Young}

\begin{document}

\title{The Cool ISM in Elliptical Galaxies. II.  Gas Content in the Volume -
Limited Sample and Results from the Combined Elliptical and Lenticular Surveys}

\author{Gary A. Welch}
\affil{Saint Mary's University}
\affil{Department of Astronomy and Physics}
\affil{Halifax, Nova Scotia B3H 3C3}
\affil{Canada}
\email{gwelch@ap.smu.ca}
\and
\author{Leslie J. Sage}
\affil{University of Maryland}
\affil{Department of Astronomy}
\affil{College Park, Maryland 20742}
\affil{USA}    
\email{lsage@astro.umd.edu}
\and
\author{Lisa M. Young}
\affil{New Mexico Institute of Mining and Technology}
\affil{Department of Physics}
\affil{Socorro, New Mexico, and}
\affil{Adjunct Astronomer at the National Radio Astronomy Observatory}
\email{lyoung@physics.nmt.edu}
    
\begin{abstract} We report new observations 
of atomic and molecular gas in a volume limited sample of 
elliptical galaxies.  Combining the elliptical sample with an earlier and 
similar lenticular one, we show that cool gas detection rates are very similar 
among low luminosity E and SO galaxies but are much higher among luminous 
S0s.  Using the combined 
sample we revisit the correlation between cool gas mass and blue 
luminosity which emerged from our lenticular survey, finding strong support 
for previous claims that the molecular gas in ellipticals and 
lenticulars has different origins.  Unexpectedly, however, and contrary to 
earlier claims, the same is {\it not true} for atomic gas.  We speculate that 
both the AGN feedback and merger paradigms  
might offer explanations for differences in detection rates, and 
might also point towards an understanding of why the two gas phases could 
follow different evolutionary paths in Es and S0s.  
Finally we present a new and puzzling discovery concerning the global mix of 
atomic and molecular gas in early type galaxies.  Atomic gas comprises a 
greater fraction of the cool ISM in more gas rich galaxies, a trend which can 
be plausibly explained.  The puzzle is that galaxies tend to cluster around 
molecular-to-atomic gas mass ratios near either 0.05 or 0.5.

\end{abstract}

\keywords{galaxies: elliptical and lenticular, cD - galaxies:evolution -
galaxies: ISM }

\section{INTRODUCTION}
With this paper we conclude our surveys of atomic and molecular gas in 
volume limited samples of elliptical and lenticular galaxies (S0s: \citet{ws03},
\citet{sag06}; Es: \citet{sag07}, Paper I).  The prime 
motivation for our work has been to address the long-standing mystery of why 
those kinds of galaxies typically have much less cool gas than their 
stars have returned over the last 10 Gyr.  We have accepted the 
penalty of long integration times in order to finesse possibly serious 
biases in earlier studies - towards optically luminous 
galaxies and/or those already known or suspected to contain large amounts of 
gas - and also in order to probe for cool gas at mass limits far below those 
implied by our current understanding of stellar evolution.

\section{OBSERVATIONS AND DATA REDUCTION}

\subsection{Data from the GBT and IRAM 30m Telescopes}

The properties of the sample have been described in 
Paper I.  The 22 HI spectra presented here were recorded at the 
NRAO Robert C. Byrd Green Bank Telescope (GBT)\footnote{The National Radio 
Astronomy Observatory is a facility of the National Science Foundation operated 
under cooperative agreement by Associated Universities Inc.} in April 2007.  
At the frequency of the 21cm hyperfine transition the GBT has 
FWHM=8.7$\arcmin$. Sensitivity limits  were designed to provide 
5$\sigma$ detections, or upper limits, of 0.02M$_e$, where M$_e$ is the mass of 
gas returned within a 10 Gyr old galaxy according to \citet{fg76}.  
Observations were conducted in seven 2-9 hour sessions between 21 and 28 April. 
Each session included flux calibration on either 3C 48 or 3C 286 and a scan of 
a spiral galaxy with strong HI emission.  Calibration observations were 
omitted during two sessions which immediately followed another 
program using the same backend 
setup.  The observing unit was a standard on-off sequence in which 150 seconds 
each were spent on the source and nearby sky.  Off positions were located 
either one-half degree east or west of the galaxy to avoid spurious emission 
from objects known from the NASA Extragalactic Database to be at a similar 
redshift. 
Data was reduced with GBTIDL following standard procedures.  Occasional bad 
scans were deleted and noise spikes removed. The baseline on each summed, 
smoothed spectrum was 
defined by a window from 2000 to 4000 km s$^{-1}$ wide, except 1000 km s$^{-1}$ 
wide in the case of UGCA 298.  The baseline window was centered on the galaxy 
optical velocity whenever possible, but was shifted to avoid Galactic 
foreground emission for low redshift objects.  The line window (Column 2 in 
Table 1) was chosen to include visible emission, otherwise it was centered on 
the systemic velocity and its width set equal to twice the measured stellar 
velocity dispersion.  First- or second-order 
polynomial fits were generally found to be satisfactory, although 4th order 
fits were made in the case of NGC 584, 636, 1172 and 4125.  Spectra were binned 
to 5.12 km s$^{-1}$ resolution.     

In June 2007 the IRAM 30m telescope on Pico Veleta, Spain was used to search 24 
galaxies for CO emission in the J=1-0 and J=2-1 transitions.  The telescope has 
FWHM=21$\arcsec$ and 11$\arcsec$, respectively, at the frequencies of those 
transitions.  Single pointed observations were made in all cases.  Sensitivity 
requirements on the J=(1-0) transition were derived using the same criterion as 
for the GBT sessions, and 
observing procedure followed the same pattern as in previous visits 
to the 30m telescope \citep{ws03}.  CLASS was used for data 
reduction, and standard procedures explained in our earlier papers were 
followed.  A linear baseline was subtracted from each summed scan, and line 
windows were defined in the same way as for the HI data.  Final spectra 
were binned to resolutions of 10.4 km s$^{-1}$, except 13 km s$^{-1}$ in the 
case of the CO(1-0) spectra of NGC 821 and NGC 4308.   

Figure 1 shows the final, baseline-subtracted spectra.  We have included a few 
CO spectra from Paper I in order to facilitate comparison with the new HI 
observations.  A summary of the measurements made from the new data is 
presented in Table 1. 

\subsection{The Impact of Differing Telescope Beam Sizes}     
  
The beam sizes of the HI and CO observations are very different (9$\arcmin$ and 
21$\arcsec$ for HI and CO(J=1-0), respectively).  Naturally, then, these data 
are not appropriate 
for comparisons of local HI and H$_2$ surface densities.  We believe, however, 
that they can be used to determine total gas masses with accuracies sufficient 
for the purposes of this investigation.  HI emission in local elliptical and 
lenticular galaxies has been mapped with the VLA and WSRT by \citet{sag06}, 
\citet{mor06}, and \citet{oos07}.  HI maps of 4 low luminosity 
ellipticals have been published by \citet{la87}.  The detected gas is usually 
several arcminutes (several tens of kpc) in diameter.  In the case of very 
gas rich systems or 
systems with close companion galaxies (our selection criteria generally exclude 
the latter) the atomic gas extends up to 10$\arcmin$ in diameter.  
Inspecting maps published in the above works indicates that the GBT (median 
beam diameter $\sim$50 kpc for our E and S0 galaxies) would see most of the 
emission in most cases, especially in low luminosity galaxies, although 
quantifying that impression is difficult.  
The Arecibo telescope (median beam diameter $\sim$15 kpc for our samples) has 
been used by ourselves  
\citep{sag06} to measure M(HI) in 6 galaxies, and by several previous workers 
whose results we have included.  It is possible that Arecibo has 
missed appreciable emission in a few galaxies, but we lack interferometer 
observations to quantify the situation.  
   
Turning to the molecular phase, the regions sampled by the 30m 
telescope have median diameters 
of 1.8 and 1.7 kpc for E and S0 galaxies, respectively.  As part of our 
lenticular survey \citep{ws03} unpublished CO observations at several 
positions across five galaxies (NGC3607, NGC4150, NGC4310, NGC4460, NGC5866) 
were made with the 30m Telescope.  It was found that the central pointing 
accounts for about half (median value 57\%) of the total emission seen at all 
positions.  That result is consistent with comparisons by one of us (Young) of 
30m fluxes with 
interferometer maps \citep{young02,young05,young08} of CO emission in a number 
of E and S0 galaxies.  Consequently we believe that the values 
of M(H$_2$) presented in the present work can be taken to represent the total 
molecular gas mass to within a factor of $\lesssim$2.

Thus, even though the beam sizes for HI and CO are quite different, they are 
reasonably well suited to the goal of detecting all, or nearly all, of the 
atomic and molecular emission.  The atomic gas is certainly present on much 
larger physical scales than the molecular gas, but this is compensated to some 
degree by corresponding differences in beam size.  Current heterodyne arrays 
such as HARP on the JCMT and HERA on the IRAM 30m telescope could 
be used improve our understanding of beam size effects by mapping the strongest 
sources in the CO(3-2) and CO(2-1) lines, respectively.  At present, though, we 
have no evidence for a luminosity trend in the relative extent of atomic and 
molecular gas, and do not anticipate systematic biases introduced by the 
differences in beam size.   

\section{RESULTS}

\subsection{Status of our Search for Cool Gas}

The present survey, like our earlier one of lenticular galaxies, differs from 
previous work in two ways.  First, the samples are 
volume limited, and are not biased towards objects already known or suspected 
to be gas rich.  While Malmquist bias is almost certainly still present, we 
have reduced it as much as possible given the state of our knowledge of 
nearby early type field galaxies embodied in the Nearby Galaxies Catalog 
\citep{tul88} and the Third 
Reference Catalogue of Bright Galaxies (RC3, \citet{RC3}).  Second, current understanding of 
stellar evolution is used to fix sensitivity limits, mandating long integration 
times for low luminosity galaxies.  Our goal in both surveys 
has been to collect in every case, either from the 
literature or from new telescope sessions, observations of molecular and atomic 
gas sufficiently sensitive to detect $\sim$1 percent of the gas predicted to 
be returned by stars during the past 10 Gyr.  

One galaxy, Haro 20, still lacks CO 
observations to our knowledge, while only NGC 4627 has apparently 
not been searched for atomic hydrogen.  We have not attempted to derive an 
upper limit on the HI content of Maffei 1 from the early, insensitive 
observation of that galaxy \citep{sp71}.  Because the 
selection criteria for the present work do not exclude nearby companions 
(unlike our earlier study of lenticulars) there are a few cases where the low
spatial resolution of single-dish observations at 21 cm hinders the attribution 
of the detected atomic gas to individual galaxies - an important problem when 
attempting to understand the fate of internally recycled material.  Cases of 
severe confusion include NGC 3226 \citep{huc94}, where nearby NGC 3227 
(type Sa) is also in the telescope beam.  Likewise, an unknown fraction of the 
atomic gas found near NGC 7464 \citep{ls94} must be attributed to neighboring 
NGC 7465.  We therefore treat NGC 3226 and NGC 7464 as HI non-detections 
despite the fact that HI has been clearly seen towards both of them.  We also 
exclude those two galaxies when making statistical comparisons with our 
lenticular galaxy sample.   Finally, we note that NGC 7468, which was briefly 
discussed in Paper I, continues to stand out for its large quantity of atomic 
gas.              
  
\subsection{Comparisons with Previous Studies}

We have located published HI and CO observations of 13 and 8 galaxies, 
respectively, in the present survey.   The majority of the HI overlap 
comprises objects fainter than M(B) = -19, some with early, insensitive 
observations.  The present HI measurements are consistent with 
published ones in all instances.  Improvements in detector technology, and our 
observational practice of scaling the detection limits with luminosity 
whenever possible, has resulted new detections or much lower limits on gas 
content in those cases.   

The new CO data are consistent with literature values for the  
majority of galaxies previously observed; only two cases of disagreement 
deserve comment. \citet{wch95} report an IRAM 30m Telescope detection of 
NGC 4125, finding I(1-0) = 2.5 K km s$^{-1}$ (without a stated uncertainty) 
over a 400 km s$^{-1}$ emission feature.  We do not detect that galaxy, 
measuring I(1-0) = 0.74$\pm$ 0.44 K km s$^{-1}$ by integrating over a similar  
velocity range.  The other case, NGC 4278, has been observed by \citet{com07}, 
who report detecting emission at both J=1-0 and J=2-1 at roughly 5-sigma using 
the same telescope but with more sensitive data than ours.  We do not 
confidently detect NGC 4278 in either transition, although the two 
measurements of I(1-0) are consistent at the combined 1-sigma level.  Very 
recently \citet{cro10} reports that the Plateau de Bure Interferometer has 
failed to detect J=1-0 emission from NGC 4278, setting an upper limit 
equivalent to 0.49 K km s$^{-1}$ at the 30m Telescope, i.e. roughly 1.6 sigma 
below our weak detection.  It does not appear, therefore, that the J=1-0 
transition in NGC 4278 has yet been reliably measured.  Our upper limit on the 
J=2-1 transition for NGC 4278 is approximately one-third of the 4-sigma 
measurement published by Combes et al.

\subsection{Cool Gas Detection Rates}

\subsubsection{Global Detection Rates among E and S0 Galaxies}

From the volume limited lenticular and elliptical samples (i.e. \citet{ws03}, 
\citet{sag07} and the present work) we have distilled 
subsets classified as E(27 galaxies), E/S0(7), S0(21), S0/Sa(8) or 
uncertain(10), using classifications 
from the Carnegie Atlas \citep{san94} when available, otherwise from the RC3.  
To reduce the effects of different selection criteria we omit the two 
elliptical sample members with companions (see above); all 
types listed as uncertain are likewise excluded.  All further references to 
morphological type issues in this paper will be to the remaining subsets unless 
otherwise indicated.  We emphasize that they encompass our separate 
lenticular and elliptical surveys.  Detection rates based on those subsets   
are compared in Table 2 to values calculated by \citet{knp99} from published 
sources, which are biased to varying degrees towards FIR emission or other 
signs of abnormality.  We caution that the rates for E/S0 galaxies are 
based on few objects.  The tabulated rates for ellipticals are consistent with 
rates derived from Table 3 after filtering by Carnegie/RC3 type.   

\subsubsubsection{Atomic Hydrogen}

It is surprising that our surveys detect atomic hydrogen more frequently than 
previous ones, since \citet{knp99} finds that large IRAS fluxes correlate with 
increased HI detection rates among early type galaxies.  That may simply be 
due, however, to our efforts to achieve greater sensitivity among fainter 
galaxies.  We find both HI and CO more frequently among lower luminosity 
galaxies (see below) whereas Table 2 lumps together all luminosities.

\subsubsubsection{Carbon Monoxide}

Given the bias of previous studies, the fact that we find a somewhat lower CO 
detection 
rate among ellipticals is consistent with the strong correlation between CO and 
FIR flux in early type galaxies \citep{knp99}.  Our detection rate of 
26\% is essentially the same as the those reported by \citet{kr96} from  
elliptical samples having IRAS 100 micron fluxes of 1.5 Jy or less 
(their Table 5).  We caution against ascribing much significance to that
agreement, however, partly because of the luminosity dependence of detection 
rates.  Also, although we omit "uncertain" classifications, the remaining 
uncertainties in Carnegie/RC3 types could significantly alter the rates in 
Table 2.   Adding all six E/S0 galaxies to type E, for example, would increase 
HI and CO detection rates to 27\% and 64\%, respectively.

The present observations are consistent with the result in 
Paper I that the 30m Telescope more frequently detects CO(2-1) among 
lenticulars than ellipticals, which was derived by comparing members of 
our original lenticular and elliptical samples.  We focus now on types 
E, E/S0 and S0 as defined above, and on cases where either or both of the 
J=1-0 and J=2-1 transitions have been detected at 3$\sigma$ or higher, and 
thereby find 2-1 detection rates of 10/10 galaxies = 100\% among S0s, 
3/3 galaxies = 100\% for type E/S0, and 3/5 galaxies = 60\% among ellipticals.  
The statistics of small numbers suggests caution when extrapolating 
those results.  
Furthermore, selection effects are present at some level, because the 2-1 
spectra from our elliptical survey are a bit noisier than spectra in the same 
transition from the lenticular study, thereby lowering detections among 
ellipticals - recall that our sensitivity requirements are translated into 
noise limits on J=1-0 spectra.  

We point out in Paper I that lower 
2-1 detection rates could indicate that ellipticals contain cooler and/or 
less centrally concentrated molecular gas than S0s.  That conclusion, 
though, is not supported by Figure 2, an update of Figure 2 in \citet{ws03}, 
which compares the intensities in the two transitions. 

\subsubsection{The Luminosity Dependence of Detection Rates}

Ignoring possible luminosity differences, the results in Table 2 support 
the long-held opinion that cool gas is more difficult to find in elliptical 
galaxies than in lenticulars.  Accounting for luminosity reveals the more 
nuanced perspective 
shown in Figure 3.  Cool gas is indeed much more likely to be found in 
lenticulars than in ellipticals, but only when comparing luminous galaxies.  
In contrast, low luminosity objects are very likely to contain detectable 
amounts of cool gas, be they type E or S0, which is consistent with previous 
HI \citep{ls84} and CO \citep{lee91} surveys.  The puzzles, then, are why 
low luminosity objects are easier to detect, and why gas is found more 
frequently among luminous early type galaxies with robust disks.  We now 
consider possible explanations.

\subsubsection{Explanations for Variations in Detection Rates}

The trends shown in Figure 3 are unlikely to reflect environmental 
effects because our selection criteria exclude galaxies within clusters.  
Likewise our target sensitivities, which scale 
with absolute luminosity, cannot produce different detection frequencies for 
samples of similarly luminous galaxies.  Early discussions of internal 
processes which might
favor accumulation of cool gas in lower luminosity systems (\citet{fg76}, 
\citet{ls84}) were based on the reasoning that red giant ejecta would, if well 
mixed, be heated to only modest temperatures because of the small stellar 
velocity dispersion.  The cooling time of such material might therefore be 
short enough to allow most of it to form dense clouds which would be more 
difficult for the occasional supernova to push out of the galaxy.  
Efforts to 
follow the evolution of red giant outflows \citep{par08} have focussed 
on the net effects of energy transfer between the ejected gas and a hot 
ambient medium without incorporating supernova events.  The growth of a central 
complex of cool gas and dust remains generally plausible.  More significant for 
the question of luminosity effects, though, is that Parriott \& Bregman do not 
find that ejecta from slower moving red giants (i.e. denizens of less massive 
galaxies) are able to cool more efficiently.  Another possibility, that 
increased rotational support among low luminosity systems could be linked to 
higher detection rates, is not supported by the work of \citet{brm97}.   

We have speculated in Paper I about the role of AGN feedback in the 
evolution of the cool ISM.  A variety of recent theoretical and observational 
work (e.g. \citet{bow06,kau08,cat09,korm09}) has 
raised the possibility that  
the AGN feedback paradigm, which emerged from efforts to address the 
cooling flow problem in X-ray galaxy clusters, might be extended to 
individual galaxy scales.  Although many details remain to be worked out, the 
emerging picture is that cooling at the center of a hot halo stimulates AGN 
activity.  The resulting outflow reheats the halo, cutting off the 
supply of fuel to the active nucleus and presumably helping destroy clouds of 
atomic or molecular gas throughtout the galaxy.  The presence of a massive hot 
halo, then, decreases the likelihood of finding cool gas.  An explanation for 
the luminosity trend in Figure 3 emerges naturally from such a picture because 
of the well known scaling relation between the luminosities of X-ray halos and 
of the stars in the host galaxy (e.g. \citet{can87,osu01}).  Without 
enough hot gas to promote AGN feedback, a low luminostiy E or S0 galaxy would 
be more likely to retain cool gas.

How might a disk component give rise to a higher detection rate?  One 
possibility, that supernova explosions are less 
effective at removing cool gas in flatter galaxies, is not supported by 
the work of \citet{dec98}.  The AGN feedback paradigm may offer a promising
line of attack, because S0 galaxies are significantly dimmer in X-rays than 
ellipticals of similar luminosity \citep{es95}.  Therefore, even bright 
lenticulars perhaps lack the 
massive hot halos needed to transfer energy to the cool ISM.  Furthermore, and 
regardless of galaxy luminosity, gas 
returned within a high angular momentum disk environment might be 
more likely to settle into an extended, dense sheet able to survive heating 
by surrounding X-ray gas.

It is widely accepted that early type galaxies have formed through a series 
of mergers.  Does the merger paradigm identify what processes might reduce 
detection rates among more luminous galaxies?  Objects resembling 
field ellipticals emerge in \emph{ab initio} simulations of 
galaxy formation \citep{nie10}.  Although most 
of the final mass is in place by z$\sim$1 small amounts continue to 
fall in thereafter, consistent with the high frequency of  
disturbance in our combined sample (see below).

A plausible explanation for Figure 3 emerges from simulations (e.g. 
\citet{naab03,gon05,jess05,naab06a,naab06b,kan07}) aimed at 
identifying the cause of the well-known dichotomy of ellipticals, namely that 
luminous galaxies are usually rotate slowly and have boxy isophotes whereas 
faint ones tend to be rapid rotators with 
disky isophotes.  It seems that boxy galaxies are the common 
outcome of so-called dry (i.e. nearly gas-free, with M$_{gas}$/M$_{stars} 
\lesssim$ 0.1) 
mergers, while (wet) mergers of gas rich disks generate  
objects resembling elongated, rotationally supported ellipticals or 
lenticulars.  AGN feedback does not play an important role in generating the 
dichotomy \citep{kan07,naab07}.  Wet mergers probably spark bursts 
of star 
formation, yet since stars form very inefficiently the resulting galaxy would 
likely contain a significant amount of gas.  In the merger 
paradigm, then, the luminosity trend in Figure 3 could arise because more 
mergers are needed to build luminous galaxies, but each merger reduces 
the gas content of the resulting object.  In this picture, mergers between 
a few luminous gas rich disks could produce modern luminous lenticulars, which 
would then be expected to contain more gas than their elliptical counterparts. 
Interestingly, \citet{bois10} find that the outcomes of wet merger 
simulations are strongly resolution-dependent, and such events might after all 
be able to produce slowly rotating elliptical galaxies.  One might then 
ascribe low detection rates among luminous ellipticals to stellar feedback 
acting on the cool ISM.   

In summary, we believe that both the currently developing paradigms of AGN 
feedback and the formation of early type galaxies in mergers offer useful 
insights into the causes of the detection frequencies shown in Figure 3.

\subsection{Observational Evidence on the Origin of the Cool Gas in Early
Type Galaxies}

\subsubsection{Cool Gas Masses}

One of the most striking results from our earlier lenticular survey (e.g 
Figure 12 of \citet{sag06}) is a cutoff of the form 
log[M(obs)]$\sim$ 0.2$\times$log[L$_B$] in the total mass of cool gas present, 
in which 
M(obs)=1.4[M(HI)+M(H$_2$)] is the observed mass of atomic and molecular gas 
and L$_B$ is the total 
blue luminosity.  Subsequent work has not clearly confirmed that feature.  
A similar cutoff was suggested in our preliminary report on  
the elliptical sample (Figure 2 of Paper I), but is not obvious when the full 
sample is considered (Figure 4a).  On the other hand, an upper cutoff appears 
to be present in the combined sample (Figure 4b).  That impression is greatly 
enhanced, however, by two galaxies at the low luminosity end of the plot.  
The current situation, then, is that the cutoff is uncomfortably 
sample-dependent.  Sensitive observations of additional low luminosity 
galaxies will help settle the question.

We now turn attention to the entire mass range covered by the data.
Early work on the distributions of M(HI) (\citet{knapp85}, \citet{war86}) and, 
separately, of M(H$_2$) \citep{lee91} in generally FIR-biased 
samples has uncovered significant differences between E and S0 galaxies, which 
presumably arise because the two phases originate in different ways in these 
galaxy types.  Does the picture change when comparing volume limited samples?  
In the spirit of previous investigators we seek to answer that question using 
the Kaplan-Meier cumulative distribution function (CDF), which is appropriate 
for samples containing upper limits \citep{fn85}.  The value of 
CDF(X) is the expected fraction of the sample in which the value of some 
parameter x is less than X.  We have carried out the calculations 
using two independent software packages: ASURV version 1.3 \citep{if90} and R 
\citep{rp09}, finding essentially the same results.  Tests such as log-rank or   
Kolmogorov-Smirnov are used to compute the likelihood that two CDFs are 
drawn from the same parent population - the R package 
employs log-rank and the Peto-Prentice modification of the Wilcoxon 
test on two censored data sets (\citet{fn85} and references therein).   

Figure 5 compares CDFs generated by R for E and S0 galaxies.  They are derived 
from mass estimates alone (left panels) and from masses scaled by the 
predictions of stellar mass return (M(pre), \citet{cio91}, right panels); the 
scaling is equivalent to dividing by L$_B$.  In light of earlier work some 
results are unexpected, and independent of scaling.  The CDFs 
of M(HI) (middle 2 plots), which overlap in places, are likely to have been 
drawn from the same parent populations with probabilities of 15-20 percent,
depending on the method used to compare the two functions.  Thus we do not  
rule out the proposition that E and S0 galaxies have {\it the same} cumulative 
mass distributions of atomic hydrogen.  That result might in fact be reasonable   
if we are correct in speculating that much of the HI in both galaxy types 
has fallen in from a surrounding reservoir.  One might then 
expect the mass of the reservoir and its central galaxy to be at least 
roughly proportional regardless of optical morphology.

On the other hand, 
the distributions of M(H$_2$) are clearly quite different, especially among 
higher masses.  The probablilites that the two samples come from the same 
parent population is 2-4 percent.  We therefore support the work of 
\citet{lee91} which points toward different evolutionary histories for the 
molecular ISM components in E and S0 galaxies.    

If only the total cool ISM is considered (bottom 2 panels in Figure 5) 
then the CDFs of E and S0 types are again rather similar, which probably 
reflects the fact that typically most of the cool ISM is atomic gas.  We find 
probabilities of 8-26 percent that the two cool gas distributions come from the 
same parent population.

\subsubsection{Molecular to Atomic Gas Mass Ratios}

We have previously (\citet{sag06}, \citet{sag07}) reported tentative evidence 
that the mix of atomic and molecular phases in galaxies with increasing amounts 
of cool gas, be they ellipticals or lenticulars, tends to shift in favour of HI.  
Adding more data makes the trend, shown in Figure 6, more convincing.  Partly 
because of our decision to use Carnegie/RC3 morphological types, earlier 
indications of an offset between ellipticals and lenticulars has disappeared.  
The paucity of data continues to hinder attempts to compare the distributions 
of molecular to atomic gas mass ratios in Es and S0s.  \citet{lee91} find no 
reliable difference between the two distributions, while an analysis similar 
to that described above returns probabilities of $\sim$40 percent that our 
samples of M(H$_2$)/M(HI) have been drawn from the same parent population.  
More detections are needed before any meaningful difference appears.

Perhaps the most striking aspect of Figure 6 is the presence of two clumps 
separated by roughly an order of magnitude along 
both axes.  We refer to the clump centered near M(obs)/M(pre)=0.1, 
M(H$_2$)/M(HI)=0.05 as gas rich , and 
to the one centred at roughly (0.005, 0.5) as gas poor.  
Overplotted curves of constant M(H$_2$)/L$_B$ and M(HI)/L$_B$ 
reveal that the separation is mostly caused by differences in HI content.  
(The outstandingly H$_2$-rich outlier near the top of Figure 6 is 
NGC 3607 whose striking central dusty and CO-rich disk(s?) presents a 
fascinating challenge for ideas of cool ISM origins (\citet{ws03}, 
\citet{lau05}).  The reader will note that the 7 HI non-detections are 
contained in the gas poor group while all 3 CO non-detections are members of 
the gas rich clump.  We do not believe that fact can be ascribed to 
significantly different HI and CO sensitivities because achieving similar 
sensitivities in the two phases has been a goal of our observing strategy 
(Section 2).

Low luminosity galaxies are known to be relatively gas rich.  We search for a 
possible luminosity difference between clump members by arbitrarily defining 
the clumps as containing the 17 galaxies towards the top left in Figure 6 
(excluding NGC 3607) and the remaining 13 galaxies towards the bottom right, 
arriving at the memberships shown in Table 4.  The two luminosity distributions 
are shown in Figure 7.  Gas rich clump members are indeed fainter by 
the equivalent of 0.7 magnitude in the mean.  The difference, however, is 
significant at only 1.2 times the combined standard deviation of the means.    
The Kolmogorov-Smirnov test indicates a probability of 18 percent that the 
two samples are drawn from the same parent population, which we interpret as 
only weak support for the hypothesis that the two distributions differ.

\subsubsection{Speculations on the Origin of the Gas}

Differences in global kinematics and morphology motivated our suggestion  
\citep{sag06} that internal processes such as cooling 
flows produce most of the molecular gas we find, while much of the HI has 
fallen in.  Are the results described above at least consistent with those 
ideas?

An explanation for why ellipticals and lenticulars might share a common CDF for 
atomic gas but have different CDFs for molecular gas (Figure 5) can be found by 
returning to our speculation that AGN feedback might produce   
different detection rates among bright galaxies (Figure 3).  We propose that 
the efficiency of whatever process couples AGN outflow energy to the cool ISM 
is linked to galaxy type by angular momentum considerations, i.e. 
a greater tendency for returned (and therefore presumably H$_2$-rich) gas in 
lenticular galaxies to settle into highly flattened disks.  
For example, simple geometry suggests that a 
highly collimated outflow might not often impact a thin sheet of gas.  
On the other hand the morphology of the central galaxy might have little 
connection to the distribution of any infalling gas, which we postulate is 
mostly HI.  In that case the influence of AGN outflows on atomic gas would be 
similar in Es and S0s, but the outflows would be more effective at removing 
molecular gas from Es than from S0s.

Merger simulations have not yet attempted to incorporate the conversion of 
atomic gas into the molecular phase.  The merger scenario, however, 
does offer a plausible explanation for why more gas 
rich galaxies might also be richer in atomic gas.  Suppose galaxies start life 
imbedded in reservoirs of atomic gas, and that various amounts of this material 
survives subsequent mergers and blowout due to star formation episodes.  
Perhaps, then, the gas poor clump comprises galaxies which for some reason have 
failed to draw down their reservoirs, or whose reservoirs have been removed 
during successive mergers.  Capture of varying amounts of reservoir gas would 
shift such objects along paths of 
nearly constant M(H$_2$)/L$_B$ towards the lower right in Figure 6.  Objects 
which have captured most nearby gas would join the gas rich clump. It is not 
clear, though, why reservoir depletion would be the kind of all-or-nothing 
process needed to produce discrete clumps.    

We have searched for other properties besides optical luminosity which might 
offer clues to the cause of the distribution in Figure 6; references are 
identified in Table 4.  Relevant to the 
infall scenario, HI interferometer observations have been 
published for 9 galaxies, and in 8 cases the atomic 
gas shows evidence of external acquisition; the evidence is equivocal in the 
case of the ninth galaxy, NGC 7013.  Because interferometry requires 
high column densities it is perhaps not surprising that all but one galaxy 
(NGC 1052) occupy the gas rich clump.\footnote{Interestingly, 
both the HI and CO emission from the dwarf S0 NGC 404, a gas rich clump 
member, have been mapped by \citet{del04} and \citet{tay04}, respectively.  
In contrast to the atomic gas, which is 
found beyond 100$\arcsec$ from the center of the galaxy, the CO occupies only 
the inner 9$\arcsec$, consistent with our speculation that most of the 
molecular gas comes from stellar mass return.}   More sensitive HI 
observations of the members of Table 4 are clearly needed.  

Returning to optical frequencies, we find 
that all four galaxies for which published observations suggest previous 
interactions (e.g. shells, counter rotating stellar components) reside in the 
gas poor clump.  It seems 
unlikely that the small clump luminosity difference could lead to a selection 
effect, but we presently have no other explanation for that curious result.
   
In summary, published observations of Figure 6/Table 4 galaxies are consistent 
with the merger paradigm in which interactions among early type galaxies are 
common, and with the notion that capturing various amounts of 
atomic hydrogen could produce the general trend seen in that figure.  They 
do not appear, however, to point towards an explanation for its bimodal
nature.  

\section{Summary and Conclusions}

We present HI and CO observations of a volume limited sample of elliptical 
galaxies, supplementing the earlier work of \citet{sag07} on the same sample 
(Paper I), and of \citet{ws03} and \citet{sag06} on an analogous group of 
lenticulars.  The observations reported here have generally, but not always, 
strengthened the trends described in Paper I and in our earlier studies of S0s. 
 We now summarize the most significant results from those three investigations 
and the present one:

1. We do not clearly confirm our earlier finding of an upper cutoff to the 
mass of cool gas in early type galaxies (Figures 4a,b).  Settling that question 
will require additional mass estimates for atomic and molecular gas in galaxies 
with L$_B$$\lesssim$10$^9$L$_{\odot}$ (Figure 4b).

2. Supporting earlier results derived from FIR-biased samples, we find 
significantly different cumulative mass distributions for {\it molecular} gas 
in E and S0 galaxies (Figure 5).  Contrary to previous work, however, we do 
not rule out the hypothesis that the cumulative distributions of {\it atomic} 
gas are the same.     

3. More gas rich early type galaxies have lesser proportions of molecular gas 
(Figure 6).  The ratio of molecular to atomic gas mass M(H$_2$)/M(HI) varies 
by roughly two orders of magnitude, from $\sim$5 for extremely gas poor 
systems to $\sim$0.05 for the most gas rich ones.  Surprisingly, the variation 
manifests itself as clumping around quite different combinations of gas content 
and molecular mass fraction.  We are presently unable to identify a reason 
for this striking result.

4. We extend previous work which shows that cool gas is generally easier 
to find in S0s than ellipticals, by demonstrating that the effect appears 
primarily among luminous galaxies (Figure 3).  Low luminosity objects of 
either type are likely to contain detectable quantities of cool gas.  

Our surveys of E and S0 galaxies provide a general picture of cool gas in early 
type galaxies in low density environments, which is largely free of the biases 
present in earlier work.  We 
have explored for the first time the relationship between the two gas phases 
across a wide range of ISM mass, with results which challenge current ideas of 
ISM evolution.  We speculate on how some of our findings might be explained, 
finding that both the AGN feedback and merger paradigms offer attractive 
possibilities.  From an observational perspective the central and obvious 
obstacle to better understanding remains low sensitivity.  Despite generous 
grants of telescope time we have still not glimpsed the cool gas 
within even most bright ellipticals.  Many detections are weak, and the 
maps required to fully account for gas missed by single pointing observations 
are not yet available.  
Finding the gas, and charting its morphology and kinematics across a 
wide range of optical luminosity, will be the task of the next 
generation of radio telescopes.  Only by taking on that task can we 
hope to answer the fundamental question of how internal processes compete with 
external ones to shape what we see.  At present the greatest 
certainties remain the ones which motivated our surveys: Stars in E and S0 
galaxies have returned much more gas than has been found, and additional gas 
sometimes falls in from outside.    
 
This work has been supported by a Discovery Grant to Welch from the Natural 
Sciences and Engineering Research Council of Canada.  Young acknowledges 
funding by NSF AST-0507432.  We thank the referee for several comments which 
helped us to improve our presentation.
  
{\it Facilities:} \facility{IRAM:30m}, \facility{GBT}

\pagebreak

\pagebreak
\begin{deluxetable}{lccccccc}
\tablewidth{0pc}
\tabletypesize{\footnotesize}
\tablecolumns{8}
\tablecaption{Integrated Intensities}
\tablehead{
\colhead{Name}  & \colhead{Window}  & \colhead{I$_{CO}$(1-0)}  & 
\colhead{rms}  & \colhead{I$_{CO}$(2-1)}  & \colhead{rms}   & \colhead{I$_{HI}$}  & \colhead{rms}  \\
\colhead{}  & \colhead{(km s$^{-1}$)}    & \colhead{(K km s$^{-1}$)}   & \colhead{(K)}  & \colhead{(K km s$^{-1}$)} & 
\colhead{(K)}  & \colhead{(K km s$^{-1}$)}   & \colhead{(K)} \\
}
\startdata
%
NGC 584  & 1585-2019  & $1.59\pm 0.72$  & 0.0088  &  $1.57\pm 1.42$  & 0.0173   & $<0.31$          & 0.0058  \\ 
NGC 596  & 1725-2027  & $<0.61$         & 0.0097  & $4.24\pm 1.38$   & 0.0221   & \nodata          & \nodata \\
NGC 636  & 1704-2016  &  \nodata        & \nodata & \nodata          & \nodata  & $<0.26$          & 0.0059  \\
NGC 720  & 1498-1992  &  \nodata        & \nodata & \nodata          & \nodata  &  $<0.40$         & 0.0072  \\
NGC 821  & 1541-1920  & $0.59\pm 0.32$  &  0.0044 &  $<0.52$         &  0.0071  & \nodata          & \nodata \\\\         

IC 225   & 1508-1591  &  \nodata        & \nodata &  \nodata         & \nodata  & 0.037            &  0.0024 \\
NGC 1052 & 1180-1700  &  $<0.93$        &  0.0097 &  $7.72\pm 1.67$  &  0.0175  & \nodata          & \nodata \\
NGC 1172 & 1548-1790  & $0.85\pm 0.39$  & 0.0071  & $<0.48$          & 0.0086   &  $<0.18$         & 0.0047  \\
NGC 1297 & 1469-1687  & $<0.34$         & 0.0067  & $0.43\pm 0.41$   & 0.0081   & $<0.16$          & 0.0045  \\
Haro 20  & 1740-1950  & \nodata         & \nodata & \nodata          & \nodata  &  $3.76\pm 0.11$  &  0.0033 \\\\

NGC 1407 & 1493-2064  &  \nodata        & \nodata &  \nodata         & \nodata  &   $<0.31$        &  0.0050 \\
NGC 3115 DW1 & 665-731 & \nodata        & \nodata &  \nodata         & \nodata  &  $<0.040$        &  0.0021 \\
NGC 3156 & 1206-1430  &  \nodata        & \nodata & \nodata          & \nodata  &  $0.59\pm 0.06$  & 0.0016  \\
NGC 3226 & 986-1246   & $0.90\pm 0.30$  & 0.0052  & $1.02\pm 0.44$   & 0.0076   & \nodata          & \nodata \\
NGC 3377 & 534-796    & $0.32\pm 0.30$  & 0.0051  & $<0.70$          & 0.0120   & \nodata          & \nodata \\\\

NGC 3379 & 710-1112   & $<0.41$         & 0.0052  & $<0.93$          & 0.0119   &  \nodata         & \nodata \\
UGC 5955 & 1177-1321  & $0.38\pm 0.14$  & 0.0035  &  $0.45\pm 0.20$  & 0.0049   & $<0.040$         &  0.0017 \\ 
NGC 3522 & 1050-1400  & $<0.16$         & 0.0023  &  $0.34\pm 0.28$  & 0.0038   &  \nodata         & \nodata \\
IC 678   & 834-1101   & $0.28\pm 0.14$  & 0.0047  &  $1.13\pm 0.52$  & 0.0087   & $<0.040$         &  0.0017 \\
NGC 3640 & 1075-1427  & $<0.53$         & 0.0077  &  $2.04\pm 0.61$  & 0.0088   & \nodata          & \nodata \\\\

NGC 3818 & 1495-1963  & $1.48\pm 0.85$  & 0.0097  & $<0.70$          & 0.0080   & $<0.20$          &  0.0038 \\    
NGC 4033 & 1491-1743  & $<0.53$         & 0.0096  & $<0.65$          & 0.0116   &  $<0.087$        &  0.0023 \\
NGC 4125 & 1127-1585  & $0.74\pm 0.44$  & 0.0053  &  $2.30\pm 0.95$  & 0.0113   & $<0.18$          &  0.0034 \\
NGC 4239 & 848-1032   & \nodata         & \nodata & \nodata          & \nodata  & $<0.051$         &  0.0016 \\
UGC 7354 & 1489-1640  & $0.50\pm 0.17$  & 0.0041  &  $1.13\pm 0.24$  & 0.0058   & \nodata          & \nodata \\\\

NGC 4278 & 383-915    & $1.52\pm 0.63$  & 0.0066  & $<0.78$          & 0.0082   & \nodata          & \nodata \\ 
NGC 4308 & 501-677    & $0.28\pm 0.13$  & 0.0028  &  $0.40\pm 0.20$  & 0.0044   & \nodata          & \nodata \\
UGC 7767 & 1125-1337  & $<0.22$         & 0.0040  &  $1.45\pm 0.38$  & 0.0070   &  $2.17\pm 0.14$  & 0.0037  \\
NGC 4648 & 1193-1635  & \nodata         & \nodata &  \nodata         & \nodata  & $<0.20$          &  0.0037 \\  
NGC 4627 & 435-597    & $0.30\pm 0.25$  & 0.0058  &  $0.96\pm 0.31$  & 0.0072   & \nodata          & \nodata \\\\

UGCA 298 & 769-901    & $0.38\pm 0.19$  & 0.0046  &  $2.15\pm 0.36$  & 0.0089   &  $1.25\pm 0.08$  & 0.0030  \\
NGC 4697 & 1076-1406  & \nodata         & \nodata &  \nodata         & \nodata  & $<0.22$          &  0.0049 \\
NGC 4742 & 1177-1363  & \nodata         & \nodata & \nodata          & \nodata  & $0.117\pm 0.048$ & 0.0041  \\
NGC 5845 & 1199-1701  & $0.43\pm 0.38$  & 0.0041  & $<0.51$          & 0.0055   & $<0.12$          &  0.0021 \\
NGC 7464 & 1777-1972  & $0.55\pm 0.27$  & 0.0054  &  $0.72\pm 0.41$  & 0.0083   & \nodata          & \nodata \\\\

\enddata

\tablecomments{Columns contain the galaxy name, location of line window, 
and for both CO lines and HI, the integrated line intensity in
the line window and its formal standard deviation along with rms channel 
noise in the smoothed spectrum.  The temperature scales are T$_{mb}$ and 
T$_A$ for CO and HI, respectively.  All upper limits are $1\sigma$ and 
they are used whenever the formal line intensity is less than $1\sigma$.  
Observations for many of the galaxies without values were reported in Paper I.}





\end{deluxetable}

\pagebreak
\begin{deluxetable}{lccccccccc}
\tabletypesize{\footnotesize}
\tablecolumns{10}
\tablewidth{0pc}
\tablecaption{Cool Gas Detection Rates}
\tablehead{
\colhead{} & \multicolumn{4}{c}{Atomic Hydrogen} & \colhead{}
		& \multicolumn{4}{c}{Molecular Gas} \\
\cline{2-5}  \cline{7-10}
\colhead{} & \multicolumn{2}{c}{Present} & \multicolumn{2}{c}{Knapp} & \colhead{}
		& \multicolumn{2}{c}{Present} & \multicolumn{2}{c}{Knapp}\\
\colhead{Type}  & \colhead{percent}  & \colhead{N} 
		& \colhead{percent}  & \colhead{N} & \colhead{} 
		& \colhead{percent}  & \colhead{N}
		& \colhead{percent}  & \colhead{N} \\ 
}
\startdata
%
	E    &	19 &	27 &	5 &	64 & \nodata &	26 &	27 &	39 & 61 \\
	E/S0 &	67 &	6  &	17 &	23 & \nodata &	67 &	6 &	31 & 26 \\
	S0   &	57 & 	21 & 	20 & 	103 & \nodata & 62 & 	21 & 	47 & 43 \\	
\enddata
\pagebreak
\tablecomments{Columns list morphological type from \citet{san94} when available, 
otherwise from \citet{RC3}, and the 
percent of galaxies detected in our combined S0 and elliptical surveys or listed by 
\citet{knp99} out of a total of N galaxies.  Columns 2-5 concern HI detections, 
while CO detections are summarized in columns 7-10. }
\end{deluxetable}

\pagebreak
\begin{deluxetable}{lcrrcc}
\tablecolumns{6}
\tablewidth{0pc}
\tablecaption{Total Cool Gas Masses}
\tablehead{
\colhead{Name} & \colhead{Type} & \colhead{M(H$_2$)} & \colhead{M(HI)} 
& \colhead{H$_2$ reference}  & \colhead{HI reference} \\
\colhead{} & \colhead{} & \colhead{($M_{\sun}$)} & \colhead{($M_{\sun}$)} & 
\colhead{}  & \colhead{} \\  
}
\startdata
%
NGC 584  & S0 1 (3,5)     & $<5.15\times 10^7$ & $<5.95\times 10^7$    &  2  &  2  \\
NGC 596  & E0/S0 1 (disk) & $ 3.72\times 10^7$ & $ 1.45\times 10^8$    &  2  &  3  \\
NGC 636  & E1             & $<1.67\times 10^7$ & $<5.30\times 10^7$    &  1  &  2  \\
NGC 720  & E5             & $<2.40\times 10^7$ & $<5.88\times 10^7$    &  1  &  2  \\
NGC 821  & E6             & $<2.23\times 10^7$ & $<3.09\times 10^8$    &  2  &  3  \\\\
NGC 855  & [E]            & $ 7.45\times 10^5$ & $ 2.68\times 10^7$    &  4  &  5  \\
IC 225   & [E]            & $ 2.50\times 10^6$ & $<3.82\times 10^6$    &  1  &  2  \\
Maffei 1 & [gE]           & $<2.08\times 10^5$ & \nodata \tablenotemark{a}   &  1  &  6  \\
NGC 1052 & E3/S0 1(3)     & $ 3.78\times 10^7$ & $ 4.53\times 10^7$    &  2  &  5  \\
NGC 1172 & S0 1(0,3)      & $<1.70\times 10^7$ & $<2.09\times 10^7$    &  2  &  2  \\\\
NGC 1297 & S0 2/3(0)      & $<1.50\times 10^7$ & $<1.91\times 10^7$    &  2  &  2  \\
Haro 20  & [E+ (doubtful)]& \nodata            & $ 2.13\times 10^8$    &  \nodata &  2 \\
NGC 1407 & [E0]           & $<3.35\times 10^7$ & $<1.52\times 10^8$    &  1  &  2  \\
NGC 2768 & S0 1/2(6)      & $ 4.99\times 10^7$ & $ 1.98\times 10^8$    &  1  &  5  \\
NGC 3073 & [SAB0-]        & $ 7.68\times 10^6$ & $ 1.66\times 10^8$    &  1  &  7  \\\\

NGC 3115 DW 1 & [dE1, N]  & $<3.33\times 10^6$ & $<2.55\times 10^6$ & 1 &  2  \\
NGC 3156 & E5/S0 2/3(5)   & $ 3.32\times 10^7$ & $ 2.41\times 10^7$    &  8  &  2  \\
NGC 3193 & E2             & $<4.46\times 10^7$ & $<7.98\times 10^7$    &  1  &  9  \\
NGC 3226 & E2/S0 1(2)     & $ 2.14\times 10^7$ &  \nodata \tablenotemark{b}  &  2  &  3  \\
NGC 3377 & E6             & $<2.51\times 10^6$ & $<4.17\times 10^6$    &  2  &  10 \\\\
NGC 3379 & E0             & $<3.48\times 10^6$ & $<2.78\times 10^6$    &  2  &  11 \\
UGC 5955 & [E]            & $<5.28\times 10^6$ & $<3.99\times 10^6$    &  2  &  2  \\
NGC 3522 & [E]            & $<5.48\times 10^6$ & $ 1.08\times 10^8$    &  2  &  12 \\
IC 678   & [E]            & $<5.61\times 10^6$ & $<4.36\times 10^6$    &  2  &  2  \\
NGC 3605 & E5             & $<1.43\times 10^7$ & $<2.39\times 10^7$    &  1  &  10 \\\\
NGC 3608 & E1             & $<1.02\times 10^8$ & $<3.48\times 10^7$    &  8  &  12 \\
NGC 3640 & [E3]           & $ 1.85\times 10^7$ & $<7.44\times 10^7$    &  2  &  12 \\
NGC 3818 & E5             & $<6.70\times 10^7$ & $<4.30\times 10^7$    &  2  &  2  \\
NGC 4033 & S0 1(6)        & $<3.96\times 10^7$ & $<1.46\times 10^7$    &  2  &  2  \\
NGC 4125 & E6/S0 1/2(6)   & $<3.38\times 10^7$ & $<3.67\times 10^7$    &  2  &  2  \\\\
NGC 4239 & [E]            & $<9.89\times 10^6$ & $<4.93\times 10^6$    &  1  &  2  \\
UGC 7354 & [E pec (unc)]  & $ 3.73\times 10^6$ & $ 1.01\times 10^8$    &  2  &  5  \\
NGC 4278 & E1             & $<7.65\times 10^6$ & $ 2.59\times 10^8$    &  2  &  13 \\
NGC 4283 & E0             & $ 3.41\times 10^6$ & $ 4.15\times 10^7$    &  1  &  3  \\
NGC 4308 & [E (unc)]      & $<8.68\times 10^5$ & $<1.13\times 10^7$    &  2  &  12 \\\\
NGC 4494 & E1             & $ 5.59\times 10^6$ & $<9.96\times 10^6$    &  1  &  11 \\
UGC 7767 & [E]            & $ 7.20\times 10^6$ & $ 8.18\times 10^7$    &  2  &  2  \\
NGC 4648 & [E3]           & $<1.60\times 10^7$ & $<2.69\times 10^7$    &  1  &  2  \\
NGC 4627 & dE5, N         & $ 2.79\times 10^6$ &  \nodata              &  2  & \nodata \\
NGC 4636 & E0/S0 1(6)     & $<7.15\times 10^6$ & $<6.12\times 10^7$    &  1  &  14 \\\\
UGCA 298 & [E+ (unc)]     & $ 2.63\times 10^6$ & $ 1.16\times 10^7$    &  2  &  2  \\
NGC 4697 & E6             & $<2.96\times 10^7$ & $<4.22\times 10^7$    &  1  &  2  \\
NGC 4742 & E4             & $<1.40\times 10^7$ & $<2.54\times 10^7$    &  1  &  2  \\
NGC 5845 & [E (unc)]      & $<2.37\times 10^7$ & $<2.07\times 10^7$    &  2  &  2  \\
NGC 7464 & [E1 pec (unc)] & $<1.29\times 10^7$ &  \nodata \tablenotemark{c}  &  2  & 15 \\\\
NGC 7468 & [E3 pec (unc)] & $ 2.35\times 10^7$ & $ 1.59\times 10^9$    &  1  &  5  \\

\enddata
\pagebreak
\tablecomments{Columns contain galaxy name, morphological type from the Carnegie Atlas or 
RC3 (square brackets), H$_2$ mass or upper limit, 
HI mass or upper limit, source of H$_2$ value, source of HI value.  
Upper limits are $3\sigma$, and results derived by other observers have 
been corrected to the distances in Paper I.  
A CO-to-H$_2$ conversion factor of 
$2.3\times 10^{20}$ mol. cm$^{-2}$ (K km s$^{-1}$)$^{-1}$ has been used.}

\tablenotetext{a}{Maffei 1: An early HI observation \citep{sp71} is too insensitive to be useful.}
\tablenotetext{b}{NGC 3226: Published HI measurement confused by NGC 3227 (type Sa).}
\tablenotetext{c}{NGC 7464: Published HI observations confused by NGC 7565.}

\tablerefs{         (1) Paper I;  (2) present work;
(3) \citet{huc94};  (4) \citet{wch95};  (5) \citet{huc95};  
(6) \citet{sp71};  (7) \citet{irw87};  (8) \citet{com07};  
(9) \citet{wil91};   (10) \citet{kna79};  (11) \citet{brg92};
(12) \citet{ls84}  (13) \citet{rai81}; (14) \citet{kt83};
(15) \citet{spr05}.  }
\end{deluxetable}
\pagebreak
%
\begin{deluxetable}{lcclcc}
\tabletypesize{\footnotesize}
\tablecolumns{6}
\tablewidth{0pc}
\tablecaption{Figure 6 Clump Members}
\tablehead{
\multicolumn{3}{c}{Gas-Rich} & \multicolumn{3}{c}{Gas-Poor} \\
\cline{1-3}  \cline{4-6} \\
\colhead{Name} & \colhead{Type} & \colhead{Reference} & 
\colhead{Name} & \colhead{Type} & \colhead{Reference} \\
}
\startdata
%
 NGC 404   &  S0    &	1 &	NGC 596   &  E/S0	&  12  \\
 NGC 855   &  E     &	2 &	IC 225    &  E  	& \nodata \\ 
 NGC 1023  &  S0    &	3, 4 &	NGC 1052  &  E/S0	&  13  \\
 NGC 2787  &  S0/a  &	5 &	NGC 2768  &  S0		&  14, 15  \\
 NGC 3073  &  S0    &	6 &	NGC 3115  &  S0		& \nodata  \\
 NGC 3522  &  E     & \nodata & NGC 3156  &  E/S0	& \nodata  \\
 NGC 3870  &  S0    &	7 &	NGC 3384  &  S0		&  16  \\
 NGC 3941  &  S0/a  &	8 &	NGC 3412  &  S0		& \nodata  \\ 
 NGC 4203  &  S0    &	9 &	NGC 3489  &  S0/a	& \nodata  \\
 NGC 4278  &  E     &	10 &	NGC 3640  &  E 		&  17  \\
 NGC 4283  &  E     & \nodata & NGC 4026  &  S0		& \nodata  \\ 
 UGC 7767  &  E     & \nodata &	NGC 4111  &  S0		&  18  \\
 NGC 7013  &  S0/a  &	11 &	NGC 4150? &  S0/a	&  19  \\
 \nodata   & \nodata &\nodata &	NGC 4494  &  E		& \nodata  \\	
 \nodata   & \nodata &\nodata &	NGC 4880  &  E/S0	& \nodata  \\	
 \nodata   & \nodata &\nodata &	NGC 5866? &  S0		& \nodata  \\	
 \nodata   & \nodata &\nodata &	NGC 7457  &  S0		&  19  \\
\enddata
\pagebreak
\tablecomments{Broad morphological types, taken from the Carnegie Atlas 
or RC3, are listed for members of the two clumps in Figure 6.  NGC 3607, near 
the top of the plot, is excluded; question marks indicate other outliers  
which might be excluded.  References point to data suggesting past 
interactions such as extended and/or misaligned HI, optical shells, 
counterrotating stars or gas, polar rings - 1: \citet{del04}, 2: \citet{wal90}, 
3: \citet{san84}, 4: \citet{mor06}, 5: \citet{sho87}, 6: \citet{irw87}, 
7:  \citet{sag06}, 8: \citet{fis97}, 9:  \citet{vdr88}, 10:  \citet{rai81}, 
11: \citet{kna84}, 12: \citet{gou94}, 13: \citet{vng86}, 14: \citet{kim89}, 
15: \citet{ber92}, 16: \citet{whi90}, 17: \citet{pru88}, 18: \citet{bar98}, 
19: \citet{ems04} }
\end{deluxetable}

\pagebreak

\begin{figure}
\figurenum{1a}
\plotone{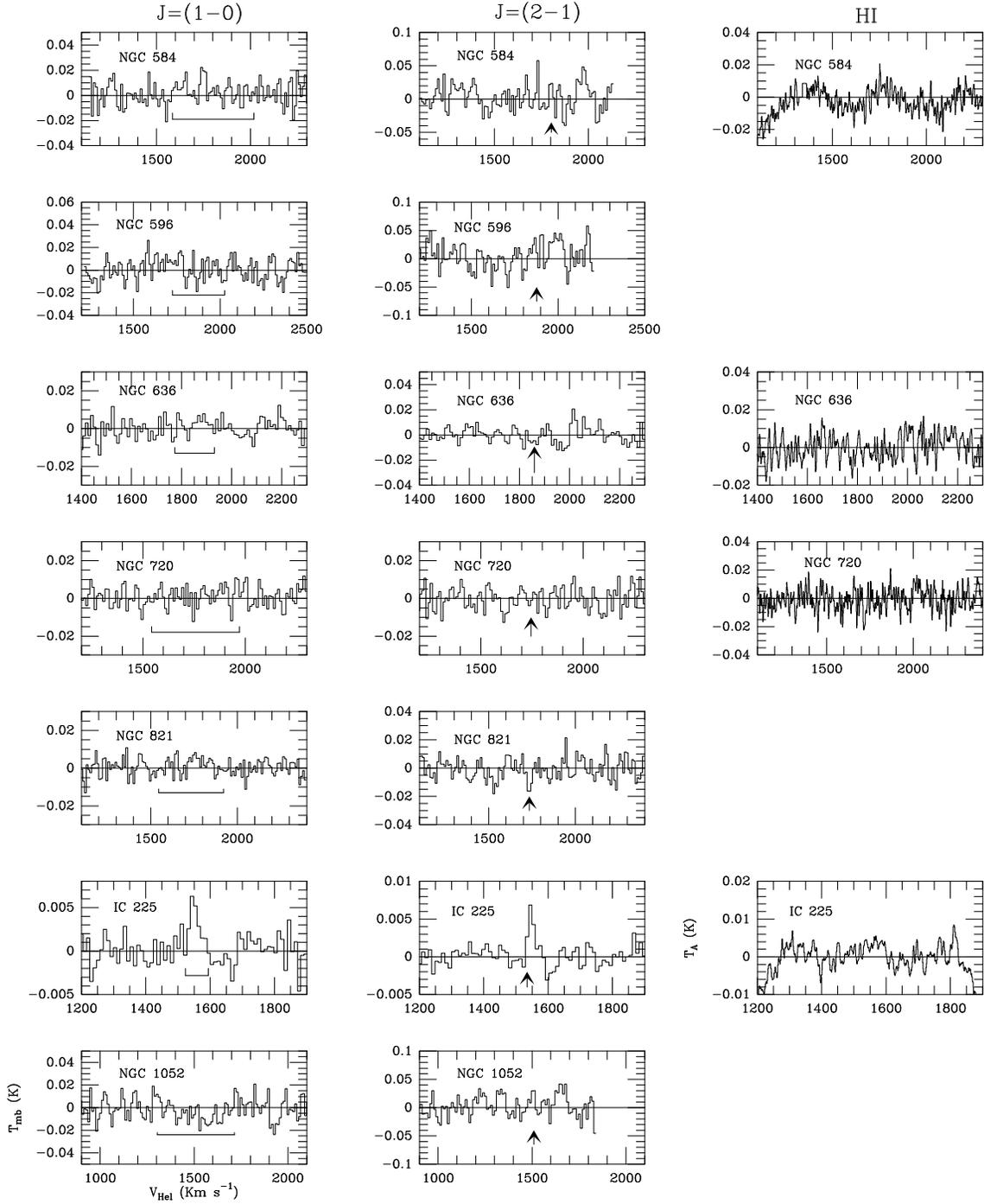}
\caption{CO and HI spectra from the IRAM 30m telescope and the GBT, 
respectively. 
Arrows indicate the optical systemic velocity from NED, and horizontal lines 
show the velocity range over which the integrated line intensity was calculated 
in all spectra. }
\end{figure}

\begin{figure}
\figurenum{1b}
\plotone{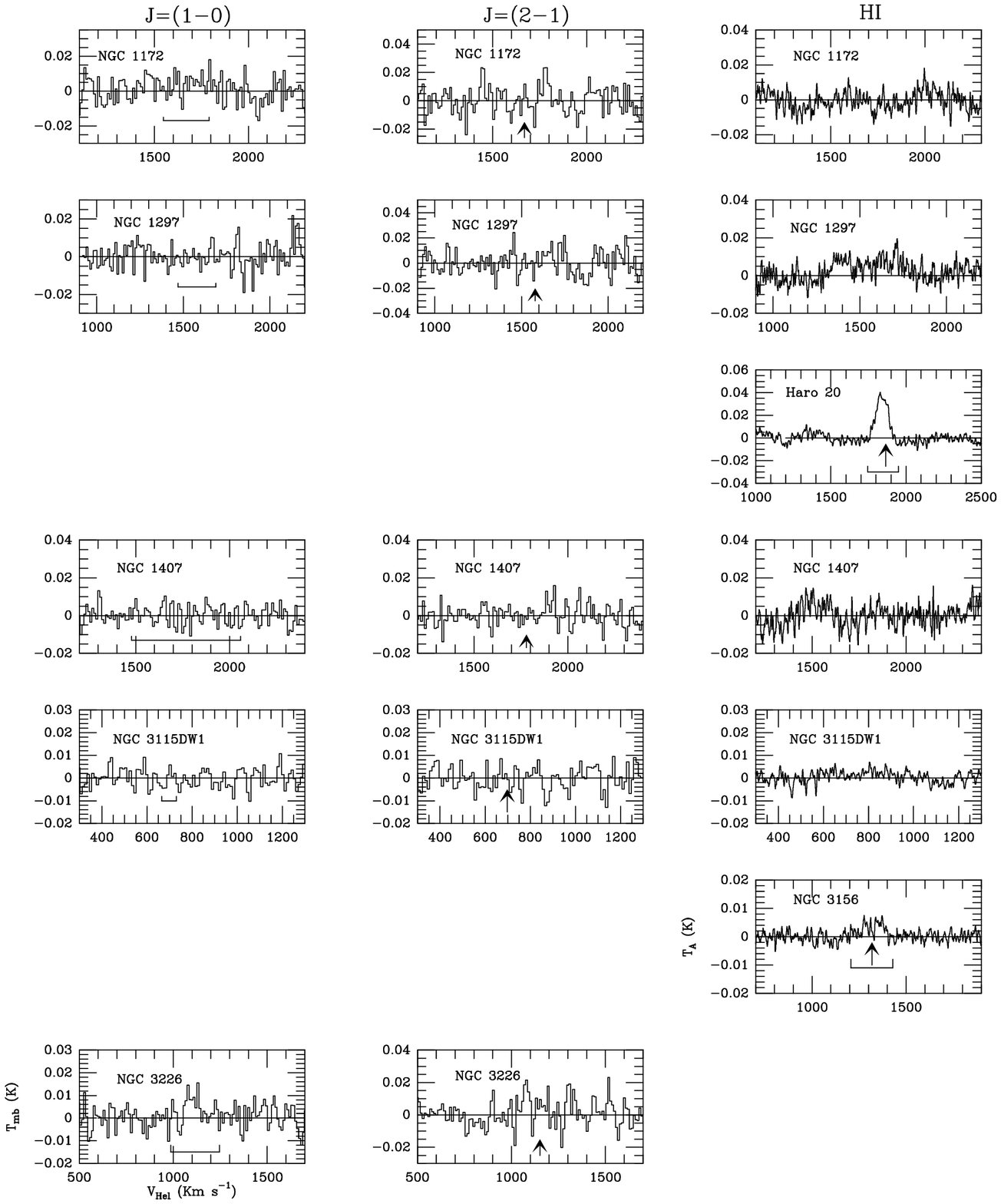}
\end{figure}

\begin{figure}
\figurenum{1c}
\plotone{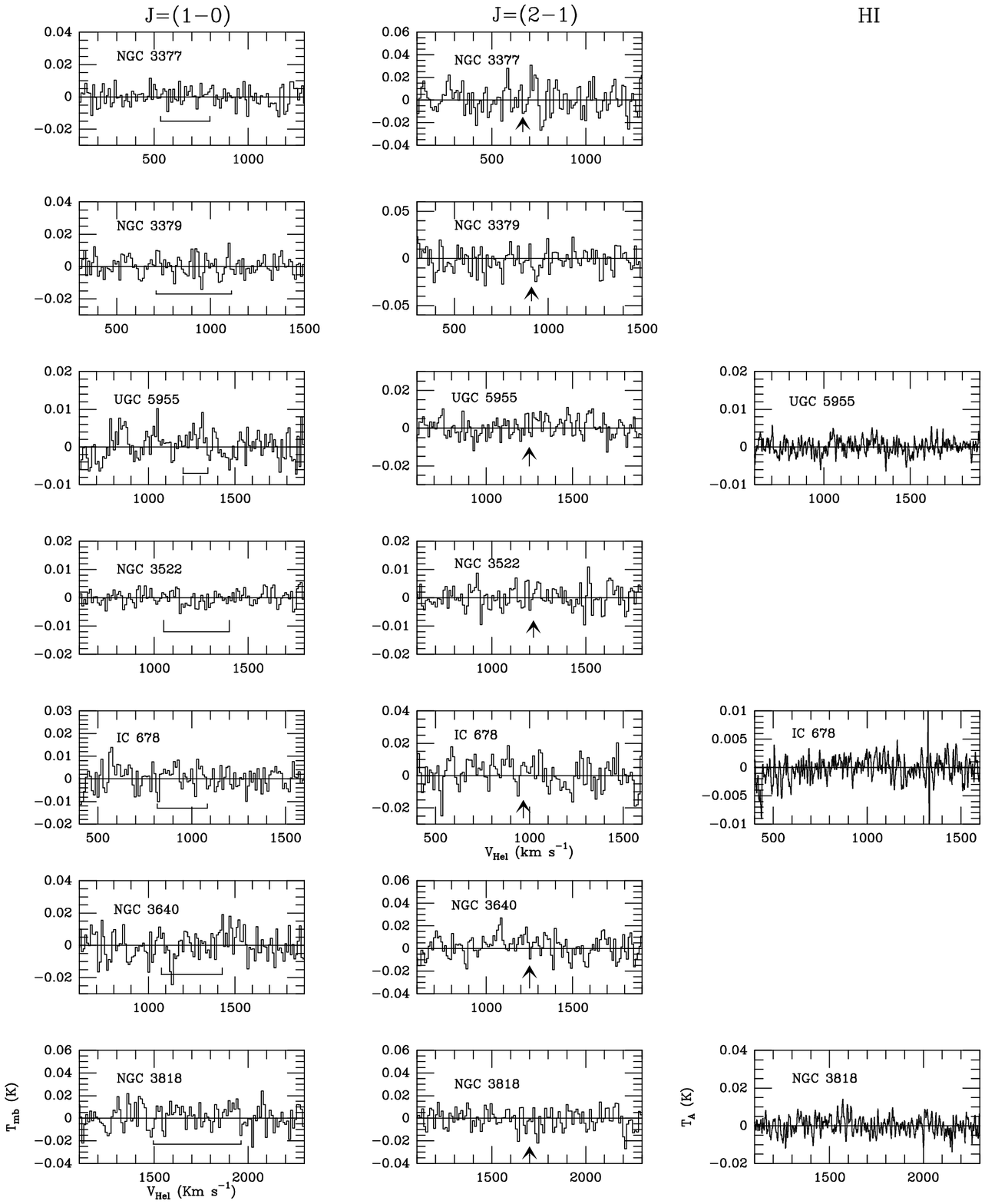}
\end{figure}

\begin{figure}
\figurenum{1d}
\plotone{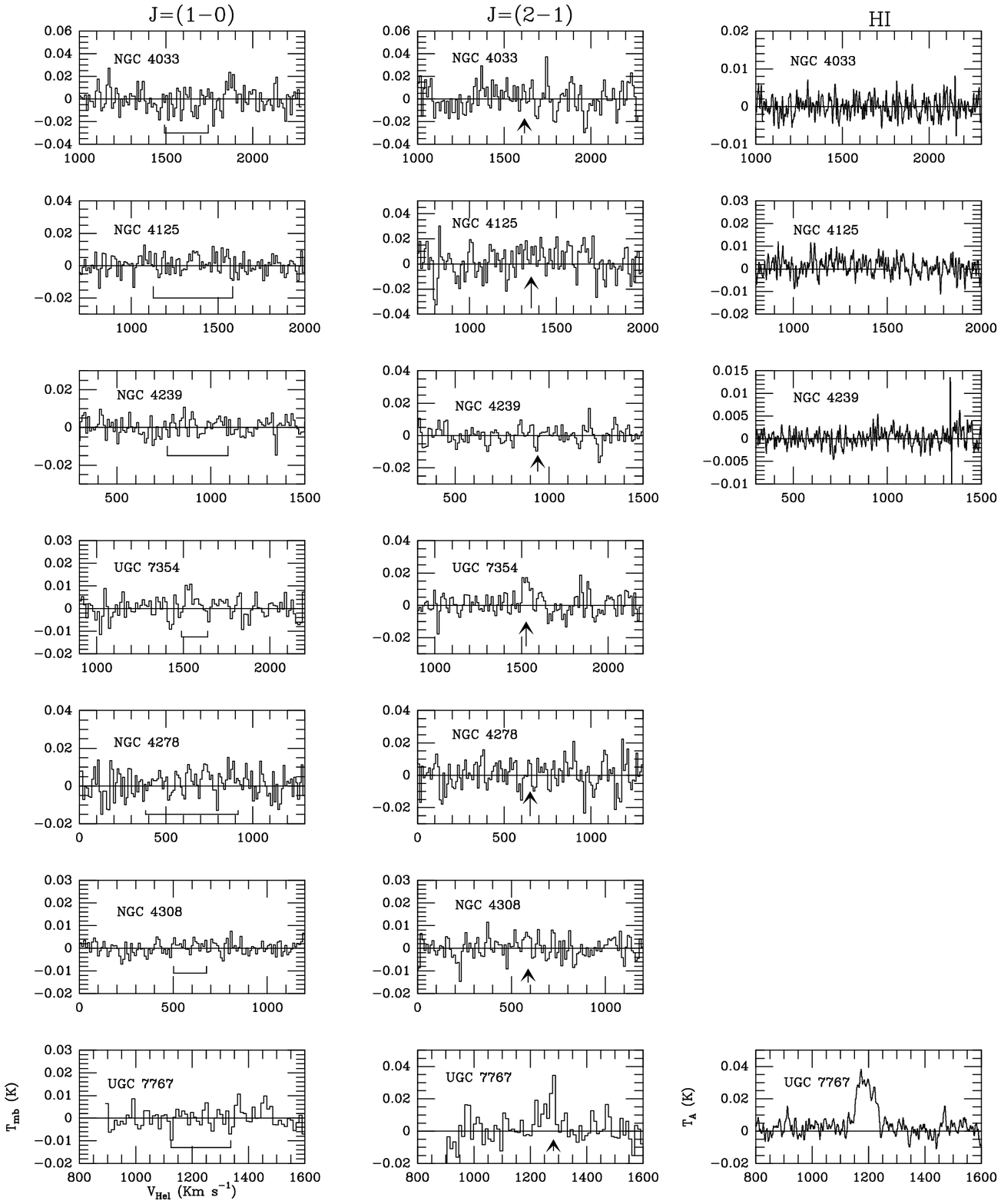}
\end{figure}

\begin{figure}
\figurenum{1e}
\plotone{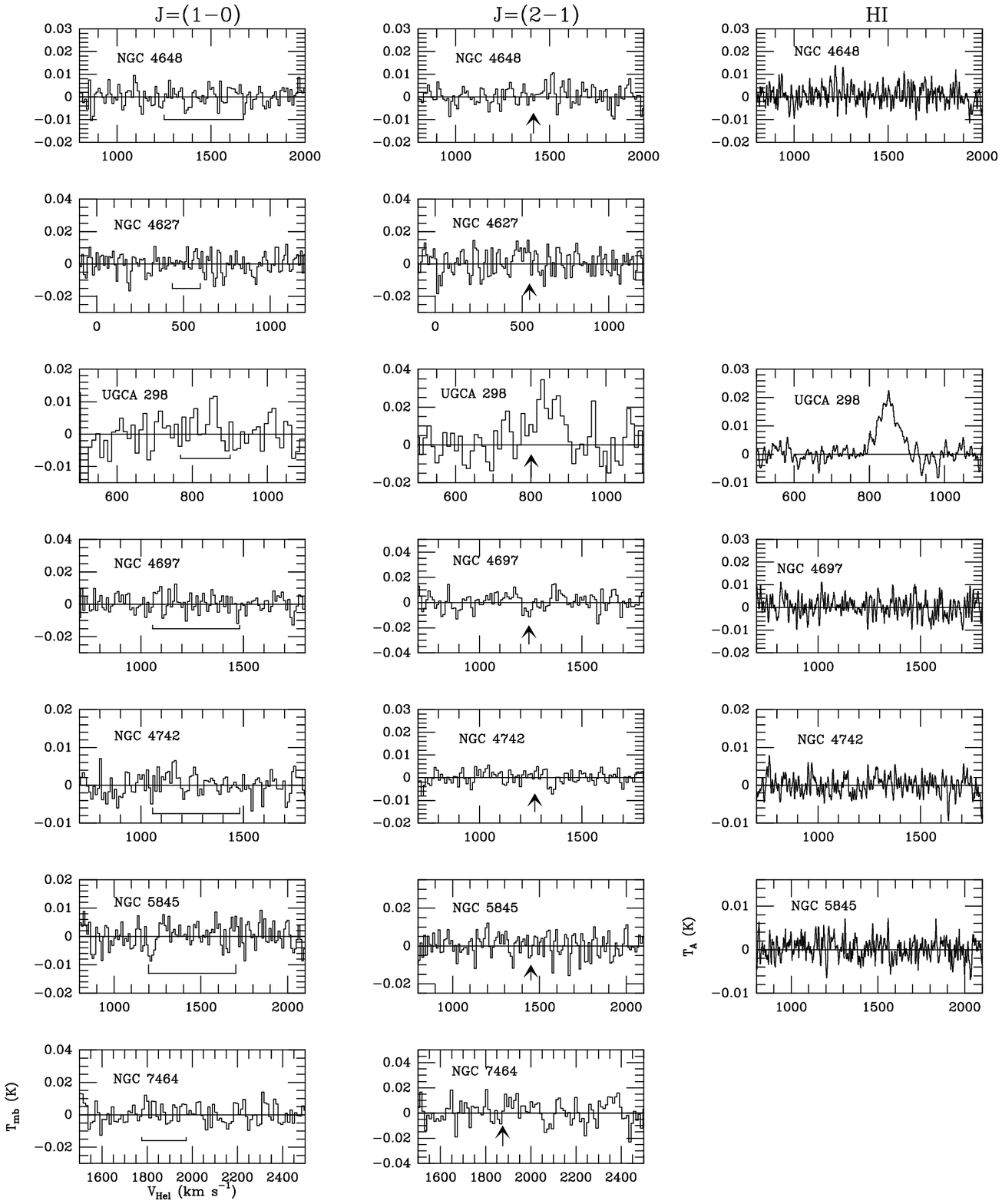}
\end{figure}
\pagebreak
\begin{figure}
\figurenum{2}
\includegraphics[scale=0.5, angle=-90]{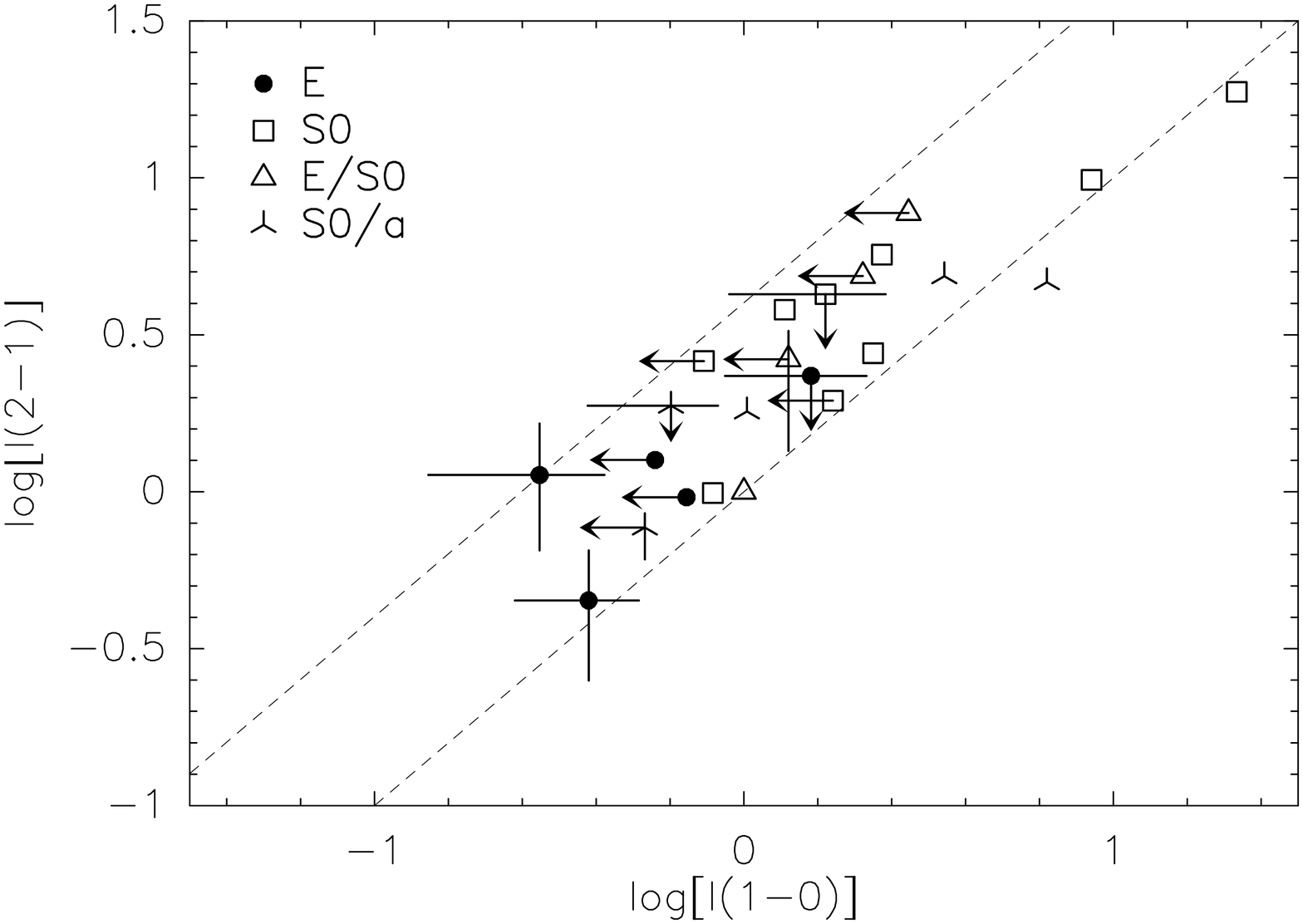}
\caption{Integrated CO intensities for galaxies detected in one or both 
indicated transitions at the 30m Telescope.  Morphological classifications are from
the Carnegie Atlas \citep{san94} or the RC3 \citep{RC3}.
Error bars have full lengths of 2$\sigma$; and are comparable to 
the size of the plotted point if not shown.  Each non-detection is  
plotted at the 3$\sigma$ value of the non-detected transition with an arrow 
extending to the 2$\sigma$ value.  Dashed lines 
indicate the relationships for point sources (top line) or those which are 
uniform across both beams.  This plot updates Figure 2 of \citet{ws03}.}
\end{figure}
\pagebreak
\begin{figure}
\figurenum{3}
\includegraphics[scale=0.5, angle=-90]{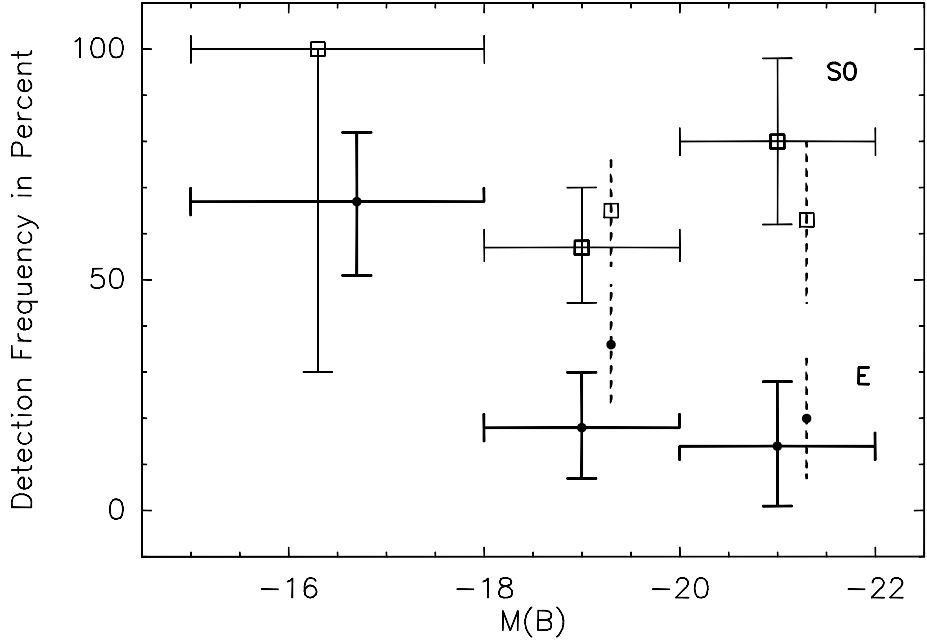}
\caption{Percent of galaxies with different absolute blue magnitudes which 
are detected in at least one of HI or CO.  Morphological types are from the Carnegie
Atlas or RC3.  Horizontal bars indicate bin sizes.  Vertical 
error bars show 1-sigma uncertainties computed assuming that detections follow 
a Binomial Distribution with detection probablility given by the ratio of the 
number of detections to the total number in the bin.  Error bars for the faintest S0 
galaxies, however, are based on the square root of the number of detections; 
there are only 2 S0 galaxies in the lowest 
luminosity bin and both are detected.  Small horizontal offsets are applied to 
separate Es and S0s in the faintest bin.  Type E/S0 is not 
shown separately because it includes only 6 galaxies.  Offset symbols with 
dashed error bars show how the E or S0 data points would change 
if all 6 E/S0 galaxies were to be assigned to one of those types; no E/S0 
galaxies would be included in the faintest bin.} 
   
\end{figure}

\pagebreak
\begin{figure}
\figurenum{4a}
\includegraphics[scale=0.5, angle=-90]{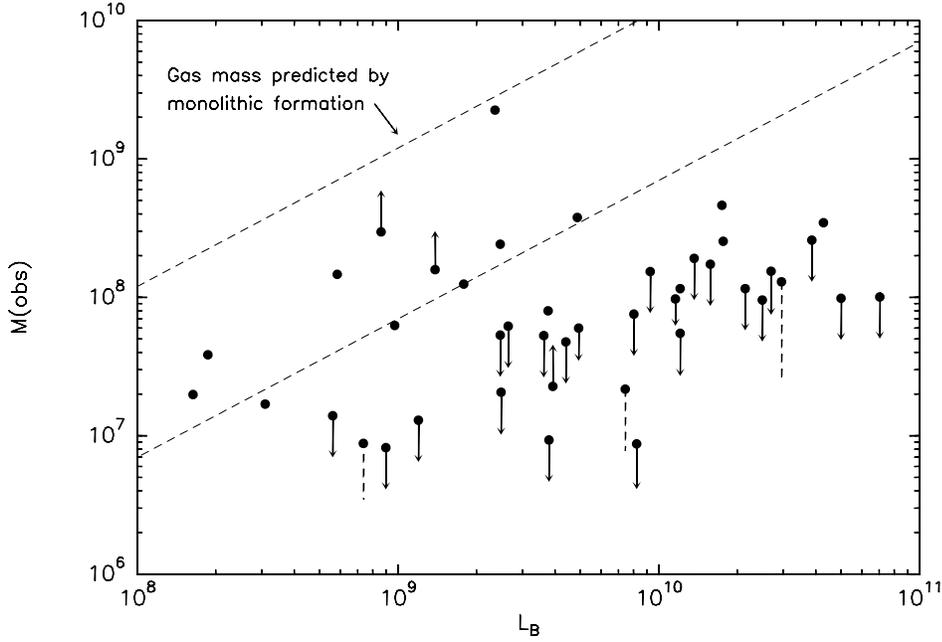}
\caption{The observed mass of cool gas versus blue luminosity among galaxies in
the elliptical sample, plotted to compare directly with results for our 
lenticular sample (Figure 12 of \citet{sag06}).  Dashed lines connect the sum 
of an observation and 3-sigma limit (point) to the observation alone.  Down 
arrows extend from the sum of 3-sigma upper limits on HI and H$_2$ masses, 
whereas up arrows identify HI detections without CO observations.  
Inclined dashed lines show the predictions of the analytical approximation 
of \citet{cio91} for gas returned in a 10 Gyr old stellar population after
the first 0.5 Gyr (top), and the estimate of \citet{fg76} for only solar type 
stars over 10 Gyr. } 
\end{figure}
\pagebreak
\begin{figure}
\figurenum{4b}
\includegraphics[scale=0.5, angle=-90]{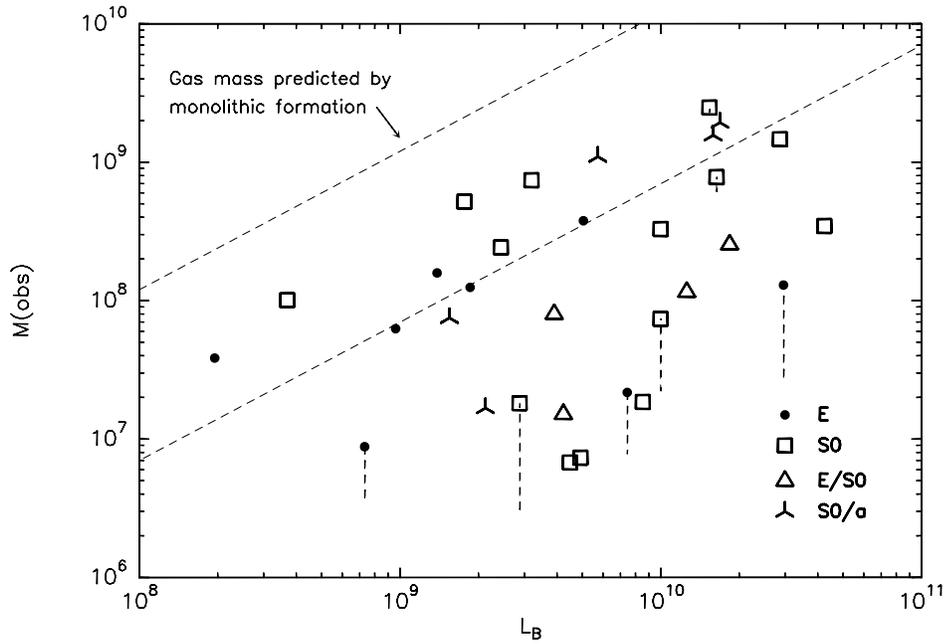}
\caption{Like Figure 4a except for galaxies in the combined E and S0 samples 
with the indicated classifications from the Carnegie Atlas or RC3, and 
which have been searched for both HI and CO and detected in at least one 
phase.  Excluded are objects having uncertain morphological type in the 
above references, and confused observations as noted in the text.  Dashed 
lines extending from data points have the same meaning as in Figure 4a. } 
\end{figure}
\pagebreak
\begin{figure}
\figurenum{5}
\epsscale{0.8}
\plotone{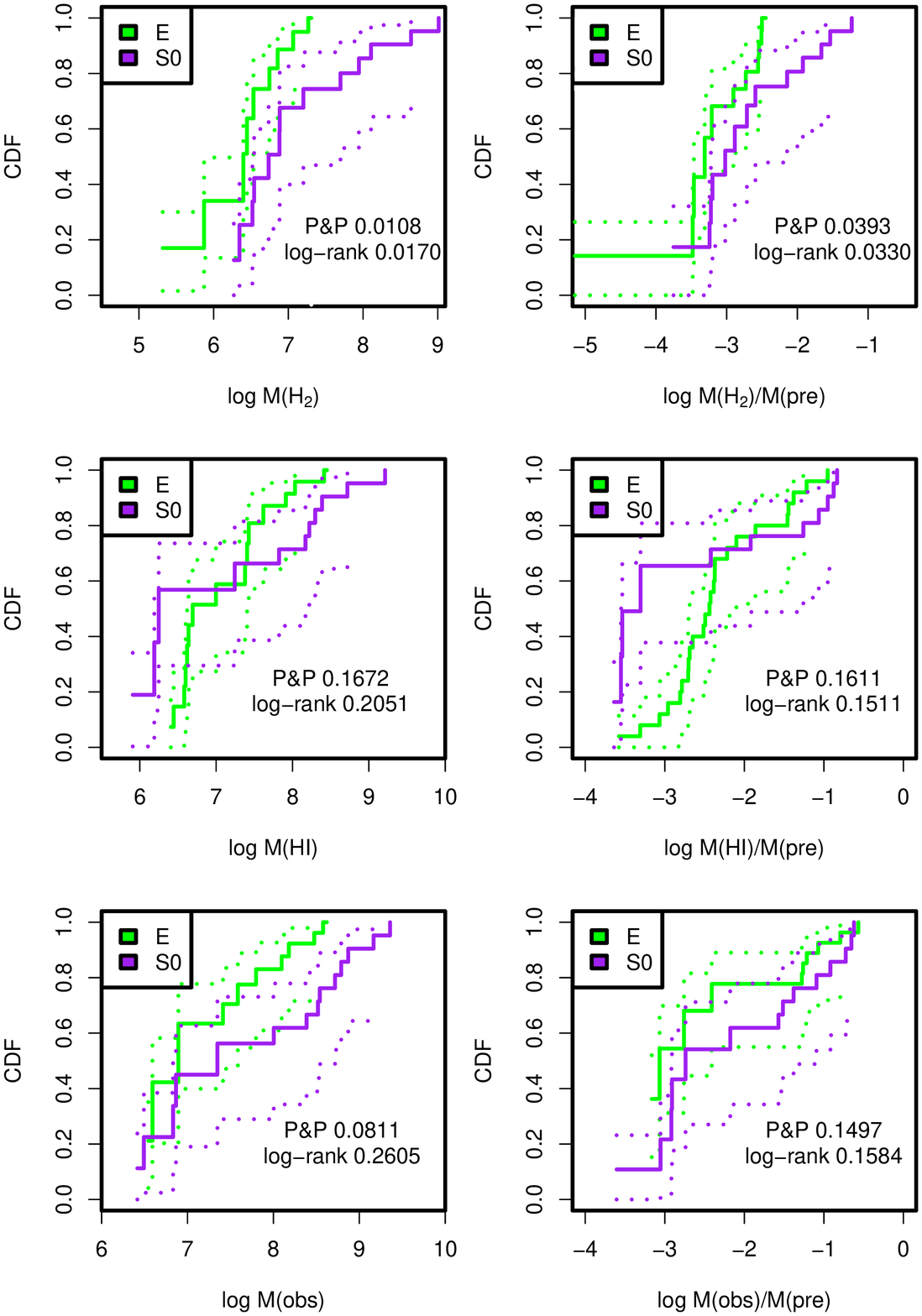}
\caption{Comparisons of the Kaplan-Meier cumulative distribution functions 
(CDF) for the atomic, molecular, and total cool gas content of galaxies in the 
combined surveys, and classified E or S0 in the Carnegie Atlas or RC3.  Dotted 
lines around each CDF trace the 95 percent pointwise confidence intervals.  
Insets give the probablilies that the two data sets have been drawn from the 
same parent population, and are dervied using the log-rank and Peto-Prentice 
tests. Masses are in solar units.} 
\end{figure}
\pagebreak
\begin{figure}
\figurenum{6}
\includegraphics[scale=0.5, angle=-90]{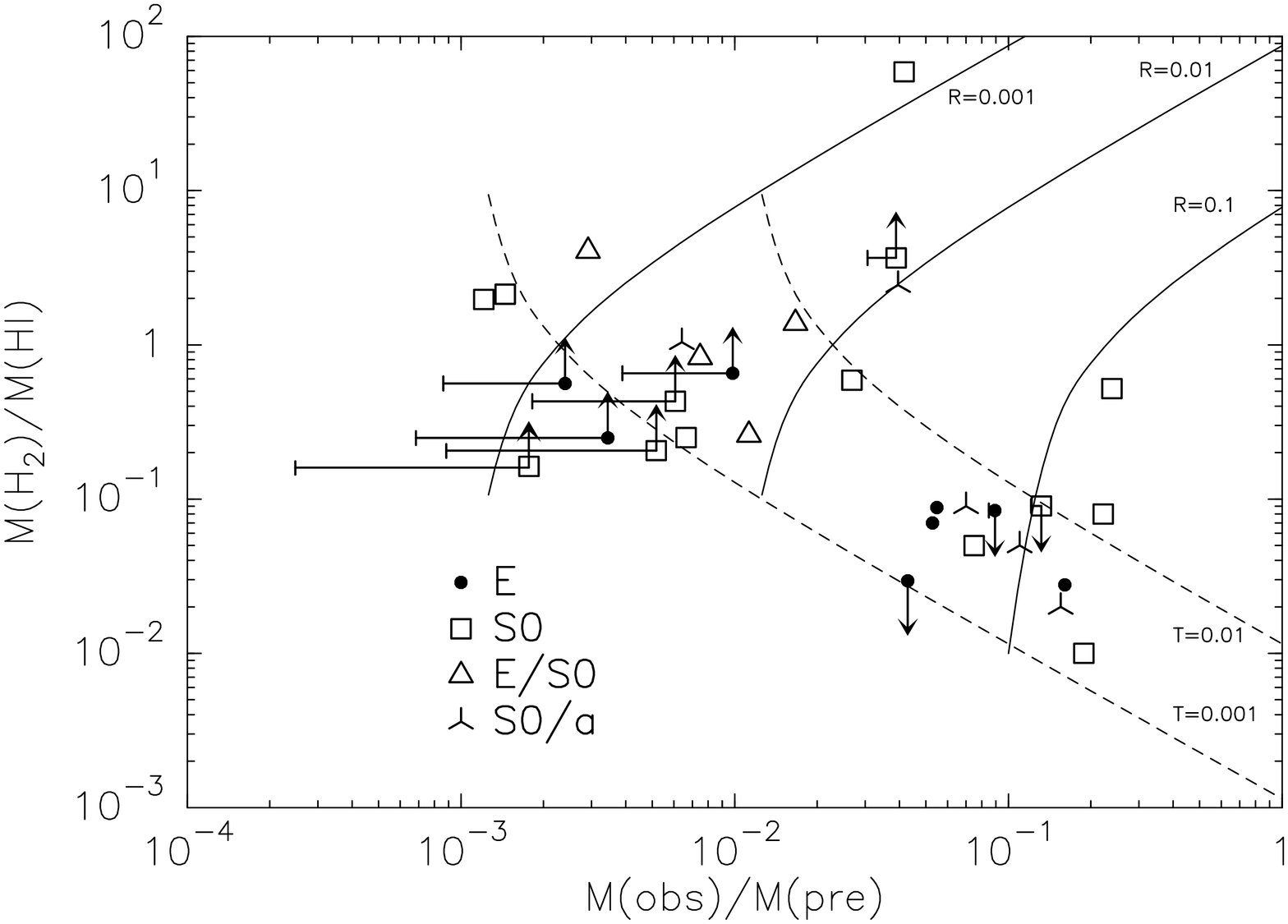}
\caption{Ratio of molecular to atomic gas mass in the combined sample, for 
galaxies classified as explained in Figure 2. Horizontal bars connect the sum 
of a detection and 3-sigma limit (point) and the detection alone.  Up and down 
arrows identify limits on M(HI) and M(H$_2$), respectively.  Loci of constant 
R=log[M(HI)/L$_B$] and T=log[M(H$_2$)/L$_B$] are shown, respectively, by solid 
and dashed curves. } 
\end{figure}
\pagebreak
\begin{figure}
\figurenum{7}
\includegraphics[scale=0.5, angle=-90]{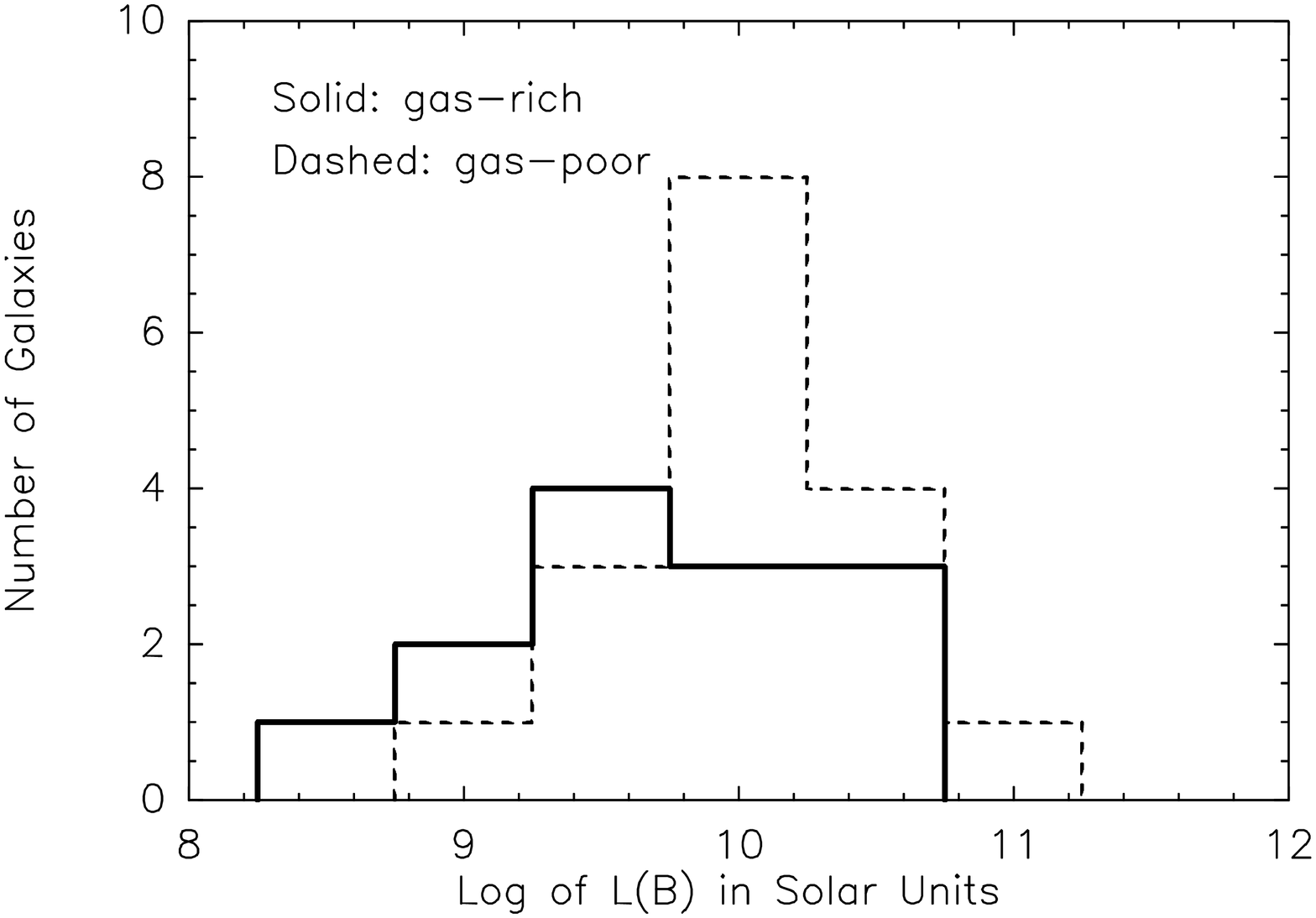}
\caption{Histogram of absolute blue luminosities for galaxies in the gas-poor 
and gas rich clumps seen in Figure 6 and identified in Table 4.} 
\end{figure}
\end{document}